\documentstyle[12pt]{article}
\begin{document}
\begin{center}
{\bf DOMAIN RIBBONS INSIDE DOMAIN WALLS\\
          AT FINITE TEMPERATURE}
\end{center}

\begin{center}
{F. A. Brito and D. Bazeia}
\end{center}

\begin{center}
{Departamento de F\'\i sica, Universidade Federal da Para\'\i ba\\
Caixa Postal 5008, 58051-970 Jo\~ao Pessoa, Para\'\i ba, Brazil}
\end{center}

\begin{center}
{Abstract}
\end{center}
In this paper we deal with defects inside defects in
systems of two scalar fields in $3+1$ dimensions. The systems we consider are
defined by potentials containing two real scalar fields, and so we are going
to investigate domain ribbons inside domain walls. After introducing some
general comments on the possibility of finding defects that support internal
structure in two specific systems, we introduce thermal effects to show how the
picture for domain walls hosting domain ribbons appears at high temperature.

\bigskip

\noindent{PACS numbers: 11.10.Lm, 11.27.+d, 98.80.Cq}

\newpage

\section{Introduction}
\label{sec:intro}

The possibility that the early universe may have experienced
symmetry breaking phase transitions resulting in the formation of defects has
provided a motivation for studies of several possible defect configurations
-- see for instance Ref.~{\cite{vsh94}}. In this route to defect formation
we can single out the case that considers the presence of defects inside
defects. This possibility was initiated in \cite{wit85}, firstly within the
context of superconducting strings, where one requires a model engendering
a $U(1)\times U(1)$ symmetry, and after in \cite{lsh85,mac88}. Other more
recent works on the same issue can be found in
\cite{mor94,mor95a,mor95b,brs96a}.

To implement the idea of finding defect inside defect, in general one
considers systems of two scalar fields, in which the first field plays the
usual role one finds in the standard route to defect formation, and
the second field enters the game via a potential that couples
it to the first field, in a way such that the system now allows for the
presence of defects inside the defect originated by the first scalar field.
This idea is usually implemented by introducing a general potential,
depending on the two scalar fields and containing several parameters that
are {\it a posteriori} tuned to allow for the presence of defects inside
defects. Despite this general picture, it was recently shown in \cite{brs96a}
that models belonging to a class of systems of two real scalar fields
\cite{bds95,bsa96,brs96b} also appear suitable to develop the idea related
to internal structure of topological defects. In this case the model is
controlled by a reduced number of parameters, and this may guide us toward
a clearer understanding of some physical aspects of the system.

The systems we shall investigate are defined with two real scalar fields in
$3+1$ dimensions, and present $Z_2\times Z_2$ symmetry that implements
spontaneous symmetry breaking in the two independent field directions.
Thus we shall be dealing with the presence of domain ribbons inside domain
walls. Although domain walls may conflict with observations,
because of wall domination within cosmological scenarios, there exist
mechanisms that allow an avoidance of wall domination. One such mechanism
relies on allowing the exact discrete symmetry to be replaced by an approximate
discrete symmetry, and this can occurs for example as a consequence
of supersymmetry breaking in supersymmetric theories \cite{mba96}. When there
are no fermions present, then the regions of higher energy density tend to
shrink, allowing closed domain walls bubbles to colapse away. The presence
of domain ribbons within walls is not expected to qualitatively change this
process, although there may be a release of boson radiation from the ribbon.

When fermions are present, however, the situation may be a little more
complicated. It has been pointed out \cite{mca95} (within the context of a
model containing no domain ribbons) that if fermions are coupled strongly
to a domain wall field, and if the fermions, which are massive in the vacuum,
become massless within the domain wall, then the Fermi gas within the wall
can contribute a degeneracy pressure which tends to stabilize the surface
area of the wall. However, the vacuum bag can flatten and fragment,
ultimately producing tiny fermionically stabilized bags of false vacuum
called ``Fermi balls''. The domain walls in this type of model
can ultimately be replaced by Fermi balls, which can be cosmologically
acceptable. It has been argued that this type of model can be obtained
from a supersymmetric domain wall model \cite{mba96}, where supersymmetry
breaking terms cause the exact wall-producing discrete symmetry to be
replaced by an approximate one.

Consider now a system accommodating fermions and domain ribbons.
A specific model is the supersymmetric system already investigated in
Ref.~{\cite{mor97}}. In this model the fermions become massless inside the
ribbons, but are massive outside the ribbons -- both inside the domain wall
and in the vacuum. Therefore there is a strong force attracting the fermions
into the ribbons from the domain wall. Thus, fermions that are initially
present within the wall may quickly be absorbed into the ribbons.
A Fermi gas of massless fermions develops within a ribbon, allowing a closed
ribbon loop to stabilize inside the wall.

Now let us again suppose that a small energy difference in the initially
degenerate vacuum states develops, so that the exact discrete symmetry
giving rise to the domain walls is replaced by an approximate discrete
symmetry. As before, we expect the space to fill with closed bags of false
vacuum, which tend to collapse. There are now two possibilities: (i) either
the typical vacuum bag will collapse away before any ribbons form within it,
or (ii) ribbons will form before the bag collapses completely. In case (i),
the end result may be the production of stable Fermi balls. In case (ii),
closed ribbon loops form within the typical vacuum bag, and the ribbons
tend to absorb the fermions from the wall, so that a stable ribbon loop
can reside in the vacuum bag. The above scenario serves to
demonstrate the possible importance of finite temperature effects,
since the dynamical pathways and intermediate states involved in cases
(i) and (ii) leading to the possible end states may depend strongly on
the difference between the critical temperatures for domain wall and
domain ribbon formation.

To explore some of the above issues, in this work we shall firstly deal
with classical features of the system introduced in \cite{brs96a} and
of another system, presented in \cite{mor95a}. This last system is defined
by a potential that is usually considered to develop the idea of introducing
internal structure to topological defects, and so we shall be also offering
a comparison between the standard procedure \cite{mor95a} and the alternative
approach introduced in \cite{brs96a}. Owing to direct interest to cosmology,
we  shall calculate the effective potential, from which we obtain the high
temperature effects in each one of these systems. The thermal effects are
obtained by following the standard works \cite{jac74,dja74,wei74},
and here we remark that the above systems are defined by potentials that
depend on two fields, and so the effective potential or, better,
the thermal effects in general introduce two critical temperatures,
driving symmetry breaking in each one of the two independent field
directions. These investigations are organized as follows. In the next
Section we introduce general considerations concerning the presence of
domain ribbon inside domain wall in two specific systems. We also investigate
classical or linear stability of the solutions we need to implement the
idea of introducing defects inside defects. In Sec.~{\ref{sec:thermal}}
we calculate the effective potential and present the high temperature
corrections to the classical potential. Here we obtain explicit expressions
for the critical temperatures in each one of the two systems under
consideration. We end the work in Sec.~{\ref{sec:comments}},
where we comment on  conclusions and possible generalizations
of the present investigations.

\section{General Considerations}
\label{sec:formal}

We are interested in systems of two real scalar fields. In this case the
general Lagrangian density is given by
\begin{equation}
{\cal L}= \frac{1}{2} \partial_{\alpha}\phi\,\partial^{\alpha}\phi +
\frac{1}{2} \partial_{\alpha}\chi\,\partial^{\alpha}\chi - 
U(\phi,\chi).
\end{equation}
Here we are using natural units, in which $\hbar=c=1$, and the metric tensor
$g^{\alpha\beta}$ is diagonal, with elements $[1,-1,-1,-1]$. $U=U(\phi,\chi)$
is the potential, in general a nonlinear function of the two fields. In the
following we shall comment on some systems of two coupled real scalar fields
described via the above Lagrangian density.

\subsection{A Class of Systems}

The class of systems of two real scalar fields that we are interested
in is defined by the following potential, as it was already stressed
in \cite{bds95,bsa96,brs96b},
\begin{equation}
\label{eq:model}
U(\phi,\chi)=\frac{1}{2} H^2_{\phi}+\frac{1}{2}H^2_{\chi},
\end{equation}
where $H=H(\phi,\chi)$ is a smooth but otherwise arbitrary function of the
fields $\phi$ and $\chi$, and $H_{\phi}=\partial H/\partial\phi$,
$H_{\chi}=\partial H/\partial\chi$. 
In this case, systems defined by the function
$H(\phi,\chi)$ present some general and very interesting properties, mainly
in $1+1$ dimensions. For instance, the second-order equations of motion for
static solutions
\begin{eqnarray}
\frac{d^2\phi}{dx^2}=
H_{\phi}H_{\phi\chi}+H_{\chi}H_{\phi\chi}\\
\frac{d^2\chi}{dx^2}=
H_{\phi}H_{\phi\chi}+H_{\chi}H_{\chi\chi}
\end{eqnarray}
are solved by field configurations satisfying the following
set of first-order differential equations
\begin{eqnarray}
\frac{d\phi}{dx}=H_{\phi},\\
\frac{d\chi}{dx}=H_{\chi}.
\end{eqnarray}
The energy is bounded from bellow, and for configurations obeying the above
first-order equations the energy gets to its minimum value, given by
\begin{equation}
E_B=H(\phi(\infty),\chi(\infty))-
H(\phi(-\infty),\chi(-\infty)).
\end{equation}
Furthermore, the set of first order differential equations can be seen as a
dynamical systems, and we can take advantage of all the mathematical tools
available to dynamical systems to deal with those equations. In particular,
one sees that the singular points of the corresponding dynamical system are
all the possible minimum energy states of the field system, and so they are
identified to the true vacuum states of the system. On the other hand, all
static configurations we can find in the above class of systems are classicaly
or linearly stable. This is interesting, and shows that perturbative quantum
corrections about static configurations can be done by just following the
standard procedure -- see, for instance, Ref.~{\cite{raj82}}.

\subsection{First System of Two Fields}

As a first example, let us focus attention on the system defined by
\begin{equation}
\label{eq:H}
H(\phi,\chi)=\lambda\left(\frac{1}{3}\phi^3-a^2\phi\right)+
\mu\phi\chi^2.
\end{equation}
In this case the potential is given by
\begin{equation}
\label{eq:bpot}
U(\phi,\chi)=\frac{1}{2}\lambda^2(\phi^2-a^2)^2+
\lambda\mu(\phi^2-a^2)\chi^2+2\mu^2\phi^2\chi^2+
\frac{1}{2}\mu^2\chi^4.
\end{equation}
This is the system already investigated in \cite{brs96a}, and here we return
to it to show that it engenders some
very specific features, unrealized in Ref.~{\cite{brs96a}}.
To see this, let us first search for the vacuum states:
They are four, two at $\chi=0$ and $\phi^2_0=a^2$, and two
at $\phi=0$ and $\chi^2_0= r a^2$. For simplicity, here we are using
$\lambda=\mu r$, and $r$ is a real, positive and dimensionless parameter.

The potential presents the following tipical forms
\begin{eqnarray}
U(\phi,0)&=&\frac{1}{2}\mu^2 r^2(\phi^2-a^2)^2,\\
U(0,\chi)&=&\frac{1}{2}\mu^2(\chi^2- r a^2)^2.
\end{eqnarray}
In this case we see that both $U(\phi,0)$ and $U(0,\chi)$ present spontaneous
symmetry breaking, and this is all we need for building defects inside
defects in the above system. In this case we can introduce meson masses
\begin{eqnarray}
m^2_{\phi}(\phi^2_0,0)&=&4\mu^2 r^2a^2,\\
m^2_{\chi}(0,\chi^2_0)&=&4\mu^2 r a^2,
\end{eqnarray}
and so $m^2_{\phi}(\phi^2_0,0)= r\,m^2_{\chi}(0,\chi^2_0)$.
On the other hand, the potential also gives
\begin{eqnarray}
U(\phi^2_0,\chi)&=&2\mu^2 a^2\chi^2+\frac{1}{2}\mu^2\chi^4,\\
U(\phi,\chi^2_0)&=&2\mu^2 r a^2\phi^2+
\frac{1}{2}\mu^2 r^2\phi^4.
\end{eqnarray}
Here we can also introduce meson masses
\begin{eqnarray}
m^2_{\phi}(\phi,\chi^2_0)&=&4\mu^2 r a^2,\\
m^2_{\chi}(\phi^2_0,\chi)&=&4\mu^2 a^2,
\end{eqnarray}
and now $m^2_{\phi}(\phi,\chi^2_0)=
r\,m^2_{\chi}(a^2,\chi)$. We also have $m^2_{\phi}(\phi^2_0,0)=
r\,m^2_{\phi}(\phi, r a^2)$ and
$m^2_{\chi}(0,\chi^2_0)= r\,m^2_{\chi}(\phi^2_0,\chi)$.
The parameter $r$ controls the meson masses, and we see
that for $ r=1$ (that is $\lambda=\mu$) the above mass values degenerate to
the single value $4\mu^2a^2$.

At this point we realize that for $r\ne 1$, that is, for
$\lambda\ne\mu$, the system presents discrete $Z_2\times Z_2$ symmetry.
The limit $r\to 1$ introduces the $Z_4$ symmetry, and this means that
the two fields have the same physical significance. This seems to pose
the question of whether will the system choose the field to host the other
field, to lead to defect inside defect. However, a closer investigation shows
that this question is in fact nonsense since the limit $r\to 1$ should be
avoided, because in this case the system of two coupled fields degenerate
into two systems of a single field each one. To see how this works explicitly,
let us rotate the $(\phi,\chi)$ plane to the $(\phi_{+},\phi_{-})$ plane,
where $\phi_{\pm}=2^{-1/2}(\chi\pm\phi)$. In this case
$H$ can be cast to the form
\begin{equation}
H(\phi_{+},\phi_{-})=2^{-1/2}\mu
[F_{ r}(\phi_{+},\phi_{-})- F_{ r}(\phi_{-},\phi_{+})],
\end{equation}
where the function $F$ is given by
\begin{equation}
F_{r}(\phi_{\pm},\phi_{\mp})=\frac{1}{2}\left(1+\frac{1}{3} r
\right)\phi^3_{\pm}- r a^2\phi_{\pm}+\frac{1}{2}
(1- r)\phi^2_{\pm}\phi_{\mp}.
\end{equation}
Here we see that the limit $ r\to 1$ decouples $\phi_{+}$ from $\phi_{-}$,
and so there is no interaction between the two fields. A lesson to learn is
then that although the original system has two independent parameters, namely
$\lambda$ and $\mu$, only their ratio $\lambda/\mu$ or $r$ is physically
relevant to the issues under consideration, and this ratio should only take
values in each one of the two distinct regions $r\in (0,1)$ or
$r\in (1,\infty)$.

Let us now focus attention on defect formation. We see
that the potential $U(\phi,0)$ presents spontaneous symmetry breaking, and so
we can have the kink solution
\begin{equation}
\phi(x)=a\tanh(\mu r\, a x)~,
\end{equation}
with energy $E_{\phi}=(4/3)\mu r\, a^3$. However, from $U(0,\chi)$ we also
have the kink solution
\begin{equation}
\chi(y)=a r^{1/2}\,\tanh(\mu
r^{1/2}\, a y)~,
\end{equation}
with energy $E_{\chi}=(4/3)\mu r^{3/2}\, a^3$. Here we
have $E_{\chi}= r^{1/2}\,E_{\phi}$, and so the parameter $r$ also controls
the energy ratio for defect formation. The picture is then the following: The
domain wall generated by the kink of one of the two fields will host the
domain ribbon generated by the kink of the other field; the host and the
nested fields are determined by the value of the single parameter $r\ne1$,
which is the same parameter that controls how mesons of the nested field
prefer to live inside or outside the domain wall.

\subsection{Second System of Two Fields}

As a second example, let us now consider the potential
\begin{eqnarray}
V(\phi,\chi)&=&\frac{1}{2}\mu^2 r^2 (\phi^2-a^2)^2+
\mu^2 (\phi^2-a^2)\chi^2+\nonumber\\
& & +\mu^2 a^2 b^2 \chi^2+\frac{1}{2}\mu^2 c^2 \chi^4.
\end{eqnarray}
Here $r$, $b$, and $c$ are real and positive parameters, and now the system
is of the form considered in \cite{mor95a}.
This potential presents the following tipical forms
\begin{eqnarray}
V(\phi,0)&=&\frac{1}{2}\mu^2 r^2(\phi^2-a^2)^2,\\
V(0,\chi)&=&\frac{1}{2}\mu^2 r^2 a^4-\mu^2 a^2(1-b^2)\chi^2+\frac{1}{2}\mu^2
c^2\chi^4.
\end{eqnarray}
We shall assume that $0< b^2<1$. In this case we see that both $V(\phi,0)$
and $V(0,\chi)$ present spontaneous symmetry breaking. However, while the
values $\phi^2_0=a^2$ and $\chi=0$ are true vacuum states, the values
$\phi=0$ and $\chi^2_0=[(1-b^2)/c^2]a^2$ are just local minima of the
potential. We make these local minima to be true vacuum states by reducing
the number of independent parameter, requiring that $r^2 c^2 =(1-b^2)^2$.
For simplicity we set $1-b^2=s^2$ and the potential is now written in terms
of two parameters, namely $r\in(0,\infty)$ and $s\in(0,1)$.
In particular, $V(0,\chi)$ can be cast to the form
\begin{equation}
V(0,\chi)=\frac{1}{2}\mu^2 \frac{s^4}{r^2} \left(\chi^2-\frac{r^2}{s^2}a^2
\right)^2,
\end{equation}
and now there are true vacuum states also at $\phi=0$ and
$\chi^2_0=(r^2/s^2)a^2$. Here we note that the potential $V(\phi,\chi)$,
written in terms of these two parameters
$r$ and $s$, does not reproduce the potential $U(\phi,\chi)$
of the former system anymore. Thus, this second system is different of the
first system in the entire region of parameters $r\in (0,\infty)$
and $s\in (0,1)$.

In this case we have the meson masses
\begin{eqnarray}
m^2_{\phi}(\phi^2_0,0)&=&4\mu^2 r^2a^2,\\
m^2_{\chi}(0, \chi^2_0)&=&4\mu^2s^2 a^2,
\end{eqnarray}
and so $m^2_{\phi}(\phi^2_0,0)=
(r^2/s^2)\,m^2_{\chi}(0, \chi^2_0)$.
On the other hand, the potential also gives
\begin{eqnarray}
V(\phi, \chi^2_0)&=&\mu^2 a^2\frac{r^2}{s^2}(1-s^2)\phi^2+\frac{1}{2}\mu^2
r^2\phi^4,\\
V(\phi^2_0,\chi)&=&\mu^2 a^2 (1-s^2) \chi^2+
\frac{1}{2}\mu^2\frac{s^4}{r^2}\chi^4,
\end{eqnarray}
and we can also introduce meson masses
\begin{eqnarray}
m^2_{\phi}(\phi,\chi^2_0)&=&2\mu^2\frac{r^2}{s^2}(1-s^2)a^2,\\
m^2_{\chi}(\phi^2_0,\chi)&=&2\mu^2 a^2 (1-s^2),
\end{eqnarray}
and now $m^2_{\phi}(\phi,\chi^2_0)=(r^2/s^2)
\,m^2_{\chi}(\phi^2_0,\chi)$. We also have $m^2_{\phi}(\phi^2_0,0)=
[2s^2/(1-s^2)]\,m^2_{\phi}(\phi, \chi^2_0)$ and
$m^2_{\chi}(0, \chi^2_0)= [2s^2/(1-s^2)]\,m^2_{\chi}(\phi^2_0,\chi)$. Here
we notice that $r$ and $s$ control the meson masses, and there are many
possible choices for these parameters. 

Let us now investigate defect formation. From the potential $V(\phi,0)$ we
can contruct the kink solution $\phi(x)=a\tanh(\mu r a x)$, which has the
same energy we have already calculated in the former system, namely
$E_{\phi}=(4/3)\mu r a^3$. In this case, however, from $V(0,\chi)$ we have
\begin{equation}
\chi(y)= (r/s) a \tanh(\mu s a y)~,
\end{equation}
and the corresponding energy is $E_{\chi}=(4/3)\mu r(r/s) a^3$. Here we get
$E_{\chi}=(r/s) E_{\phi}$, and so we can control this energy relation by just
controlling the ratio between the two parameters $r$ and $s$. 

Here the picture is richer than the one that appears in the former system,
evidently. For instance, from the above calculations we see that values at
$s=r$ in the range $(0,1)$ are interesting values. Furthermore, the value
$s^2=1/3$ is very peculiar and imposes no restriction
on $r$: This appears from the meson masses, which allow introducing the
function
\begin{equation}
g(s^2)=\frac{2s^2}{1-s^2}.
\end{equation}
This function depends only on $s^2$ and controls the ratio between meson
masses of the field to be nested inside the domain wall. However, since
$g(s^2)\le 1$ for $s^2\le 1/3$, and $g(s^2)> 1$ for $s^2>1/3$, we see that
evaporation of domain ribbons \cite{mor95a} into elementary mesons may or
may not induce back reaction on the domain ribbon, and this appears to be
controlled by the parameter $s$. As we have already shown, this is not the
case in the former model since there we have just one parameter,
and so there is no other parameter to be tuned anymore. For
$s^2=1/3$ the above function becomes unit, and the meson masses degenerate
into a single value, irrespective of the meson being inside or
outside the domain wall.

\subsection{Classical Stability}

Since we are interested in implementing the idea of introducing internal
structure to topological defects, we should also investigate
if the topological defects are classically or linearly stable. Such a
investigation seems to be important because it put forward results that
may unveil the range of parameters where perturbative quantum
corrections can be implemented standardly. 

This is the main motivation to investigate classical stability of the pairs
of solutions we have already introduced. Before doing that,
however, we recall that the defects one is dealing with comes
from kinks that appear in the corresponding $1+1$ dimensional systems, and
so the informations we are requiring can be obtained by just investigating
these $1+1$ systems. Furthermore, we already know \cite{bsa96,brs96b} that
the first system presents stable solutions. Thus, we are left with the
issue of investigating classical stability only for the second system.

This system is identified by the following potential
\begin{equation}
\label{eq:gpot}
V(\phi,\chi)=\frac{1}{2}\mu^2 r^2 (\phi^2-a^2)^2+
\mu^2\phi^2\chi^2-\mu^2 a^2 s^2\chi^2+\frac{1}{2}\mu^2 \frac{s^4}{r^2} \chi^4.
\end{equation}
As we have already shown, it presents the two pair of solutions:
\begin{eqnarray}
\phi_1(x)&=&a\tanh(\mu rax),\qquad\chi_1(x)=0,
\\
\chi_2(x)&=&(r/s)a\tanh(\mu sax),\qquad\phi_2(x)=0.
\end{eqnarray}
We consider fluctuations about each one of these two pair of solutions, in
the form $\phi(x,t)=\phi(x)+\sum_i\eta_i
\cos(w_i t)$ and $\chi(x,t)=\chi(x)+\sum_i\xi_i
\cos(w_i t)$. We procced standardly, and we get the following Schr\"odinger
operators, which respond for classical or linear stability,
\begin{equation}
S^{(1,2)}_{d}=-\frac{d^2}{dx^2}+
V^{(1,2)}_{d}(x),
\end{equation}
where $d=\phi\phi$ or $d=\chi\chi$, and 
\begin{eqnarray}
V^{(1)}_{\phi\phi}(x)&=&4\mu^2r^2a^2+
6\mu^2r^2(\phi_1^2-a^2),
\\
V^{(1)}_{\chi\chi}(x)&=&2\mu^2a^2(1-s^2)+
2\mu^2(\phi_1^2-a^2),
\\
V^{(2)}_{\phi\phi}(x)&=&2\mu^2\frac{r^2}{s^2}a^2(1-s^2)+
2\mu^2\left(\chi_2^2-\frac{r^2}{s^2}a^2\right),
\\
V^{(2)}_{\chi\chi}(x)&=&4\mu^2s^2a^2+
6\mu^2\frac{s^4}{r^2}\left(\chi_2^2-
\frac{r^2}{s^2}a^2\right).
\end{eqnarray}

The above problems were already solved in quantum mechanics.
They are identified to modified P\"oschl-Teller systems, and everything one
needs is given in Ref.~{\cite{mfe53}}. The general results can be resumed as
follows: For the first pair of solutions, that connects $(-a,0)$ to $(a,0)$
by a straight line with $\chi=0$ we have to introduce the condition
\begin{equation}
\label{eq:sta1}
2s^4+r^2s^2\le r^2,
\end{equation}
in order to ensure stability of this pair of solutions. For the second pair
of solutions, that connects $(0,-(r/s)a)$
to $(0,(r/s)a)$ by a straight line with $\phi=0$ we have to introduce the
condition
\begin{equation}
\label{eq:sta2}
2r^2+s^2\le 1.
\end{equation}
These conditions appear after investigating the minimum energy eingenvalue
of each one of the four Schr\"odinger operators just introduced.

The above results $(\ref{eq:sta1})$ and $(\ref{eq:sta2})$ show that there is
room for choosing the parameters $r$ and $s$ without changing stability of
the solutions. In particular, if one sets $s^2=1/3$, Eqs.
$(\ref{eq:sta1})$ and $(\ref{eq:sta2})$ imply that $r^2=1/3$, also.
Here we recall that the value $s^2=1/3$ was already shown to be peculiar,
since it makes the field that generates defects to be nested inside the
domain wall to have the same mass, irrespective of being inside or outside
the wall. Furthermore, if one sets $r=s$, one sees from $(\ref{eq:sta1})$ and
$(\ref{eq:sta2})$ that now one has stable solutions only in the range
$r^2=s^2\in(0,1/3]$. Recall that $r=s$ makes the energy of each one of the two
solutions we are considering to colapse into a single value. These results
are interesting and will be further considered in the next Section,
where we deal with high temperature effects.

\section{High Temperature Effects}
\label{sec:thermal}

The above investigations lead us to pictures for building defects inside
defects at zero temperature. However, to present investigations appropriate
to the standard cosmological scenario we think that we should consider
thermal effects, since one knows that the cosmic evolution occurs via
expansion and cooling. Toward this goal, let us now deal with the effective
potential, in order to investigate how the vacuum states of the system of
two coupled real scalar fields change when the high temperature corrections
are introduced.

In the following we shall first review the main steps to get to the thermal
effects in the general system of two real scalar fields. In the sequel, we
introduce the results for the specific class of systems of two scalar fields,
defined via the function $H(\phi,\chi)$. Our investigation follows
with two subsections, in which we calculate the critical temperatures for
each one of the two systems introduced in the former Section.

\subsection{General Calculations}

We follow the standard route to symmetry breaking at high temperature, as we
have already learned from the works \cite{jac74,dja74,wei74}. In this case
the one loop contributions to the effective potential can be cast to the
general form
\begin{equation}
U^1=\frac{1}{2}\int \frac{d^\nu k}{(2\pi)^\nu} \ln \det M,
\end{equation}
where the matrix $M$ is given by
\begin{equation}
M=\pmatrix{k^2+U_{\phi\phi}& U_{\phi\chi}\cr
U_{\chi\phi}&\;k^2+U_{\chi\chi}},
\end{equation}
where the derivative of the potential has to be calculated
at constant and uniform field configurations. We can rewrite
this result as
\begin{equation}
U^1=\frac{1}{2}\int \frac{d^\nu k}{(2\pi)^\nu}\,
\Bigl[\ln \left(k^2+M^2_{+}\right)+\ln \left(k^2+M^2_{-}
\right)\Bigr],
\end{equation}
where
\begin{equation}
\label{eq:m+-}
M^2_\pm=\frac{1}{2}\left(U_{\phi\phi}+U_{\chi\chi}\right)
\pm\frac{1}{2}\,\sqrt{\left(U_{\phi\phi}+U_{\chi\chi}\right)^2-
4U_{\phi\chi}U_{\chi\phi}}.
\end{equation}

To get to the thermal effects we should set
\begin{equation}
\int dk_0\rightarrow \frac{1}{2\beta}\sum_{n=-\infty}^{\infty},\qquad
k_0\rightarrow\frac{2n\pi}{\beta},\qquad \beta=\frac{1}{T}.
\end{equation}
In this case we have
\begin{equation}
U^1_\beta=\frac{1}{2\beta}\sum_i\sum_{n=-\infty}^{\infty}\int \frac{d^{\nu-1}
k}{(2\pi)^{\nu-1}}\,
\ln \left(\frac{4\pi^2 n^2}{\beta^2}+E_{M_i}^2\right),
\end{equation}
where we have set $E_{M_i}^2=\vec{k}^2+M_i^2$, with the understanding that
$M_1=M_{+}$ and $M_2=M_{-}$. 

Let us now work in the $3+1$ dimensional spacetime. In this case, after
performing summation and integration we get, taking into account only the
high temperature effects,
\begin{equation}
U^1_\beta=\frac{1}{24\beta^2}\left(M^2_{+}+M^2_{-}\right).
\end{equation}
We use the values presented in Eq.~{(\ref{eq:m+-})} to obtain
\begin{equation}
U^1_\beta=\frac{1}{24\beta^2}\left(U_{\phi\phi}+U_{\chi\chi}\right),
\end{equation}
and so we get to the final result
\begin{equation}
U_\beta=U(\phi,\chi)+\frac{1}{24\beta^2}\left(U_{\phi\phi}+
U_{\chi\chi}\right).
\end{equation}

Symmetry exists when there is no spontaneous symmetry breaking. However,
since we are dealing with two fields we
have to impose conditions for the two independent field directions, and this
will lead us to two critical temperatures. Here the results are
\begin{eqnarray}
\left(T^c_{\phi}\right)^2&=&-24\frac{\bar{U}_{\phi\phi}}
{\bar{U}_{\phi\phi\phi\phi}+\bar{U}_{\chi\chi\phi\phi}},\\
\left(T^c_{\chi}\right)^2&=&-24\frac{\bar{U}_{\chi\chi}}
{\bar{U}_{\phi\phi\chi\chi}+\bar{U}_{\chi\chi\chi\chi}},
\end{eqnarray}
where the bar over the potential indicates that after derivating the
potential we should set $\phi=0$ and $\chi=0$.

For systems defined by $H(\phi,\chi)$, the above expressions for the critical
temperatures can be written in a better form, in terms of the function $H$. In
this case we have to replace, for instance,
\begin{eqnarray}
U_{\phi\phi}&\to&
H^2_{\phi\phi}+H_{\phi}H_{\phi\phi\phi}+
H^2_{\phi\chi}+H_{\chi}H_{\phi\phi\chi},\\
U_{\chi\chi}&\to&
H^2_{\chi\chi}+H_{\chi}H_{\chi\chi\chi}+
H^2_{\phi\chi}+H_{\phi}H_{\phi\chi\chi}.
\end{eqnarray}
The other terms can be written straightforwardly. However, if one takes the
point of view that perhaps the most interesting systems are defined by
potentials that contain at most the fourth power in the fields, then we
should only consider functions $H(\phi,\chi)$ that contain at most third
power in the fields. In this case we have simpler expressions for the quartic
derivative of the potential, and they are, explicitly,
\begin{eqnarray}
U_{\phi\phi\phi\phi}&\to&
3(H^2_{\phi\phi\phi}+H^2_{\phi\phi\chi}),\\
U_{\chi\chi\chi\chi}&\to&
3(H^2_{\chi\chi\chi}+H^2_{\phi\chi\chi}),\\
U_{\phi\phi\chi\chi}&\to&
2(H^2_{\phi\phi\chi}+H^2_{\phi\chi\chi})+\nonumber\\ 
& &+H_{\phi\phi\phi}H_{\phi\chi\chi}+
H_{\chi\chi\chi}H_{\phi\phi\chi}.
\end{eqnarray}
The above results are general results, and now we focus attention on the two
systems already introduced in Sec.~{\ref{sec:formal}} to calculate explicit
expressions for their critical temperature.

\subsection{Critical Temperature in the First System}

We are interested in investigating the system defined via the function
$H(\phi,\chi)$ introduced in Eq.~{(\ref{eq:H})}. Here the critical
temperatures are given by
\begin{eqnarray}
\left(T^c_\phi\right)^2 &=&\frac{12 r^2a^2}{3 r^2+ r+2},\\
\left(T^c_\chi\right)^2 &=&\frac{12 r^2 a^2}{r(r+5)}.
\end{eqnarray}
These results show that
\begin{equation}
\left(T^c_\phi\right)^2 =\left(T^c_\chi\right)^2\;f(r),
\end{equation}
where $f(r)$ is given by
\begin{equation}
f(r)=\frac{r(r+5)}{3r^2+r+2}.
\end{equation}
We see that $f(0)=0$, $f(1)=1$, and $f(\infty)=1/3$. Furthermore, $f(r)$ is
monotonicaly increasing for $r\in (0,1)$, and monotonicaly decreasing for
$r\in(1,\infty)$, with $f(1)=1$ as its maximun value. This leads to the
result that $\left(T^c_{\phi}\right)^2$ is always lesser than
$\left(T^c_{\chi}\right)^2$, irrespective of the value of $r$. We remark
that $T^c_{\phi}$ and $T^c_{\chi}$ cannot coincide because $r\ne1$.

\subsection{Critical Temperature in the Second System}

Let us now focus attention on the second system, defined
via the potential given by Eq.~{(\ref{eq:gpot})}. Here the critical
temperatures are given by
\begin{eqnarray}
\left(\bar{T}^c_\phi\right)^2 &=&\frac{12r^2a^2}{1+3r^2},\\
\left(\bar{T}^c_\chi\right)^2 &=&\frac{12r^2 s^2 a^2}
{r^2+3 s^4}.
\end{eqnarray}
We can write
\begin{equation}
\left(\bar{T}^c_\phi\right)^2 =\left(\bar{T}^c_\chi\right)^2\;f(r,s),
\end{equation}
where the function $f(r,s)$ is given by
\begin{equation}
f(r,s)=\frac{1}{1+3r^2}\left(\frac{r^2}{s^2}+
3 s^2\right).
\end{equation}
This function presents the following two interesting possibilities of being
unit: For $r=s$, in the interval $(0,1)$, and for $s^2=1/3$, irrespective of
the value of $r$. However, from stability results of the former Section we
see that the two critical temperatures may colaspe into a single one in the
range $r^2=s^2\in(0,1/3]$.

\subsection{High Temperature Considerations}

The high temperature results obtained for the first system show that such
system breaks the symmetry firstly in one field direction, and it is only
after this symmetry breaking that the second symmetry breaking will appear.
On the other hand, we know that this system allows the presence of defects
inside defects only after the occurrence of the second symmetry breaking.
These critical temperatures are $T^c_{\chi}$ and $T^c_{\phi}$, respectively,
and so we see that there is a temperature range, that can be written via
$t=T/T^c_{\chi}$ as $\sqrt{f(r)}\le t\le1$, in which the system only supports
structureless domain walls. In the second system there is a range in
parameter space where $r^2=s^2\in(0,1/3]$ that makes the two temperatures
$\bar{T}^c_{\phi}$ and $\bar{T}^c_{\chi}$ to colapse into a single one.

Another interesting issue concerns structureless domain walls versus
domain walls that support domain ribbons. To shed some light on this, let
us first recall that $U(\phi,\chi)$ given by $(\ref{eq:bpot})$ has the
following single field limits
\begin{eqnarray}
U(\phi,0)&=&\frac{1}{2}\mu^2 r^2(\phi^2-a^2)^2~,\\
U(0,\chi)&=&\frac{1}{2}\mu^2(\chi^2- r a^2)^2~.
\end{eqnarray}
However, from the results for the critical temperature in the first model we
see that $(T^c_{\chi})^2$ is always greater than $(T^c_{\phi})^2$. Thus if
one thinks on defect formation within the cosmological scenario, it is not
unreasonable to suppose that the host domain wall is generated by the $\chi$
field. Owing to compare this to structureless domain walls we focus attention
on the (single field) system defined by $U(0,\chi)$.

We recall that $r$ is real and positive, and so we see that the above
potential $U(0,\chi)$ presents standard domain wall, structureless. For this
system the critical temperature can be written as
\begin{equation}
T_c^2=4ra^2~,
\end{equation}
and so we can get
\begin{equation}
\label{eq:compwr}
(T^c_{\chi})^2=
\left(\frac{3}{5+r}\right)\,T^2_c~.
\end{equation}
The above result compares the critical temperature for formation of
structureless domain walls in a system of just one field to the critical
temperature for formation of domain walls that support
domain ribbons in the first system of two fields. This result shows that the
presence of the second field, which responds for nesting defects inside the
wall, contributes reducing the critical temperature. The reduction depends
on $r$, and this parameter can be controlled to directly affect
the picture for defect formation within the cosmological scenario.

Another high temperature result can be introduced in the following
way: In the first system of two fields we still consider $r$ real and positive,
but now we make $\lambda=-\mu r$. This possibility was already considered
in \cite{bmo97}, but there the motivation is directly related to investigations
of an enlarged system where the symmetry $Z_2\times Z_2$ is changed
to become $Z_2\times U(1)$, with $U(1)$ implemented globally. The present
interest is however to keep the symmetry as the discrete $Z_2\times Z_2$.
In this case the potential $U(\phi,\chi)$ given by $(\ref{eq:bpot})$ changes to
\begin{equation}
{\bar U}(\phi,\chi)=\frac{1}{2}\mu^2r^2(\phi^2-a^2)^2-
\mu^2 r (\phi^2-a^2)\chi^2+2\mu^2\phi^2\chi^2+
\frac{1}{2}\mu^2\chi^4.
\end{equation}
Now it is not hard to realize that spontaneous symmetry breaking gets
implemented only from the $Z_2$ symmetry associated to the $\phi$ field.
This means that domain walls generated by this $\phi$ field cannot host
domain ribbons. Although in this new model domain ribbons cannot be nested
inside domain walls anymore, we believe that it is still interesting to
investigate how the high temperature effects enter the game in this case too.
Here we take advantage of the investigations already done to introduce the
ratio between the critical temperatures in this system and in the (single
field) system defined by ${\bar U}(\phi,0)$. This ratio is controlled by
$[3r^2/(3r^2-r+2)]^{1/2}$. It is lesser or equal to unit for
$r\in(0,2]$, and greater for $r\in(2,\infty)$, and so the
critical temperature in the system of two fields
defined by ${\bar U}(\phi,\chi)$ is lesser or equal to the critical
temperature in the system of one field defined by ${\bar U}(\phi,0)$ for
$r$ lesser or equal to $2$, and greater for $r$ greater than $2$.

\section{Comments and Conclusions}
\label{sec:comments}

In this work we have investigated the possibility of introducing defects inside
defects in systems of two real scalar fields. After presenting some general
considerations, we have investigated the high temperature thermal effects to
the classical potential. These investigations were done on two specific
systems, the first being defined by a function $H=H(\phi,\chi)$, and the
other defined by a more general potential, as occurs in models usually
considered to build defects inside defects. The basic motivation for
investigating these two systems is to provide a comparison
between the standard approach to defects inside defects, and the alternative
route recently introduced in \cite{brs96a}.

The present investigations show that systems belonging to a general class of
systems of two real scalar fields present all the features one needs to
implement the idea of nesting domain ribbons inside domain walls. These
systems are simpler because they are defined via the function $H(\phi,\chi)$,
are controlled by a reduced set of parameters, present stable configurations,
and can be extended to become supersymmetric \cite{bmo97} very naturally.
This is interesting since one keeps the underlying features of systems clearer
than the features that appear in the standard approach. As we have seen,
however, in this system there are two critical temperatures, driving symmetry
breaking in each one of the two independent field directions. The possibility
of having two distinct critical temperatures implies that the internal
structure of the domain wall cannot appear simultaneously with the domain wall
itself. This is in distinction to results in the second system, in which it
is possible to introduce a single critical temperature to drive symmetry
breaking in the two field directions simultaneously. The relation
between the two critical temperatures in the first system is 
$(T^c_{\phi})^2=[r(r+5)/(3r^2+r+2)](T^c_{\chi})^2$ and depends on $r$,
the ratio between the two parameters $\lambda$ and $\mu$ that defines
the system. In connection with issues discussed in Sec.~{\ref{sec:intro}}
we see that $r$ is directly related to distinct possibilities of productions
of (i) Fermi balls or (ii) ribbon loops. For instance, for $r\approx1$ the
system seems to favor ribbon loops instead of Fermi balls. This same $r$
also controls the relation between the critical temperatures for formation
of structureless domain walls $(T_c)$ and domain walls that support domain
ribbons $(T^c_{\chi})$. The specific relation is given by:
$T^c_{\chi}/T_c=[3/(5+r)]^{1/2}$.

The systems of two coupled real scalar fields introduced in
Sec.~{\ref{sec:formal}} can be seen as the real bosonic sector of a
supersymmetric theory \cite{bmo97}. Within this context, if we follow the
point of view of supersymmetry to implement the idea of nesting domain
ribbons inside domain walls, we can very naturally introduce fermions
into the system. Thus supersymmetry may be very naturally used to guide
investigations to more realistic models. For instance, instead of
considering the $Z_2\times Z_2$ symmetry we may use the $Z_2\times U(1)$
symmetry, and this may make the domain wall charged, or yet the
$U(1)\times U(1)$ symmetry that is the way to get to the string territory,
where the original idea of introducing internal structure to topological
defects was firstly implemented. Some investigations are connected
to ideas presented in Sec.~{\ref{sec:intro}}. Other investigations are
directly related to the recent works \cite{pet92,pet96}.
For instance, in \cite{pet96} a $Z_2\times U(1)$ surface current-carrying
domain wall model was investigated, but there the system is defined by 
a general potential of the form of our second system of two coupled fields,
and all the results are implemented numerically due to difficulties in
finding analytical solutions to the corresponding equations of motion.
These works seem to deserve further considerations, now within the alternate
way that considers systems defined via $H(\phi,\chi)$, because in this case
there are interesting general situations where we can find explicit
analytical solutions \cite{bmo97} to the equations of motion. The
several motivations \cite{pet96} for calculating internal quantities to
such domain walls broaden with the fact that they may be calculated
analytically. These and other related issues are presently under consideration.

\bigskip

We would like to thank J.R. Morris and R.F. Ribeiro for
interesting comments. DB and FAB also thank Conselho Nacional de
Desenvolvimento Cient\'\i fico e Tecnol\'ogico, CNPq, and
Coordena\c c\~ao de Apoio ao Pessoal do Ensino Superior, CAPES, for partial
suport and for a fellowship, respectively.

\end{document}